\documentclass{epl}
\usepackage{epsfig}
\usepackage{amssymb}
\usepackage{amsmath}

\newcommand{\qav}[1]{\langle #1 \rangle }

\title{Current-current correlations in hybrid superconducting and
normal metal multiterminal structures}

\shorttitle{Current-current correlations in multiterminal structures}

\author{
G. Bignon\inst{1}
\and M. Houzet\inst{2}
\and F. Pistolesi\inst{1}
\and F. W. J. Hekking\inst{1}}

\shortauthor{G. Bignon \etal}

\institute{
\inst{1}
Laboratoire de Physique et Mod\'elisation des Milieux Condens\'es,
Centre Nationale de la Recherche Scientifique and Universit\'e
Joseph Fourier, B.P. 166, 38042 Grenoble, France \\
\inst{2} Commissariat \`a l'\'Energie Atomique, DSM,
  D\'epartement de Recherche Fondamentale sur la Mati\`ere Condens\'ee, SPSMS, F-38054 Grenoble, France
}


\begin{document}

\maketitle

\begin{abstract}
We consider a hybrid system consisting of two normal metal leads
weakly connected to a superconductor.
Current-current correlations of the normal leads are studied in the tunneling limit
at subgap voltages and temperatures.
We find that only two processes contribute to the cross-correlation:
the crossed Andreev reflection (emission of electrons in different
leads) and the elastic cotunneling.
Both processes are possible due to the finite size of the
Cooper pair.
Noise measurements can thus be used to probe directly the
superconducting wave function without the drawbacks appearing in
average current measurements where the current is dominated by direct
Andreev reflection processes.
By tuning the voltages it is possible to change the sign of
the cross correlation.
Quantitative predictions are presented both in the
diffusive and ballistic regimes.
\end{abstract}

\date{\today}


\section{Introduction}

The Andreev reflection is the process of elastic transfer of
two electrons forming a Cooper pair from the superconductor
to the normal metal \cite{Andreev}.
Electrons that are emitted in the normal metal cross the normal/superconductor
interface within a distance of the order of the size of the Cooper pair.
It is thus possible for the two electrons to be transmitted in
two different normal leads, when the distance between the
contacts is comparable with the pairs' size.
This process, called crossed Andreev reflection (CAR),
can be used to probe directly the spatial structure of the
Cooper pair, but how can it be detected?
Different authors have proposed to measure the conductance
matrix in the multiterminal structure and
to extract the distance dependent contribution
of the CAR \cite{Byers,Deutscher,Loss,Falci}.
However this procedure has two main drawbacks.
First, CAR comes along with elastic cotunneling (EC)\cite{AverinNazarov},
the transfer of an electron from one normal lead to the other one
through the superconductor (conserving its spin and energy).
Second, CAR and EC contributions to the average currents are
dominated by direct Andreev reflection in each normal
lead, {\em i.e.}, by the current associated with the transfer of
two electrons in the same lead \cite{Falci}.
We will show that one can pick up the CAR or the EC contribution
directly by measuring the cross-correlation of
currents in the two normal leads.

As a matter of fact, due to the discrete nature of charges,
current fluctuates in time around its mean value.
Current noise measurement revealed a powerful probe
for electronic systems \cite{Blanter}.
In particular, cross-correlation in multiterminal superconducting
hybrid structures has deserved some attention since it was first anticipated
\cite{Datta,Martin} and then predicted
\cite{TorresMartin,TML,LMB,Shechter,Taddei,Borlin,Samuelsson,Samuelsson2}
to show a sign change with respect to normal metallic
structures.
A simplified explanation of this effect is the following.
If electrons in the two leads are emitted from one Cooper pair, one
expects a positive correlation, since both electrons appear at the
same time in each lead.
By contrast, electrons from a fermionic source due to Pauli principle arrive one by
one in each channel and they are transmitted either in one
or the other lead,
leading to a negative correlation.
Most calculations consider Y-shaped structures attached to a superconducting
lead through a single junction \cite{TorresMartin,TML,LMB,Shechter,Borlin,Samuelsson,Samuelsson2}.
This geometry is the standard one for interference in optical
experiments, but it leads to a variety of elementary processes for
charge transfer in electronic samples.
This makes the interpretation of the cross-correlation's sign more
subtle \cite{Book}.

In this paper we consider the different situation (depicted in
Fig.~\ref{layout}) where {\em each} normal arm is {\em
directly} connected to the superconductor through a tunnel interface.
We find that CAR contributes to the cross-correlation with a
positive sign, while EC contributes with a negative sign, and there are
no spurious contributions from direct Andreev reflection.
The CAR and EC contributions can be selected independently by tuning
the voltages of the normal terminals.
Cross-correlation in tunneling systems thus provides a direct way
({\em i}) to probe both CAR and EC and ({\em ii})
to measure the sign change of the correlations.

We present quantitative predictions for the dependence of the
cross-correlation spectral noise on the distance between normal arms
both in the ballistic and diffusive regimes.

\section{Effective Hamiltonian}

\begin{figure}
    \centerline{\includegraphics[scale=0.35]{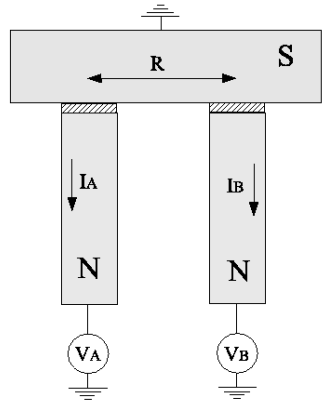}}
    \caption{Schematic picture of the three terminal device.}
    \label{layout}
\end{figure}

Let us consider a conventional Bardeen-Cooper-Schrieffer
superconductor weakly connected to two normal metal leads called
A and B (see Fig. \ref{layout}).
To describe charge transport we use the standard tunnel Hamiltonian:
$H = H_{S} + H^{A}_{N} + H^{B}_{N} + H^{A}_{T} +H^{B}_{T}$ where
$H^{\alpha}_{N}$ and $H_{S}$ describe the clean or impurity-disordered
normal lead $\alpha$ (with $\alpha= A \ \textrm{or} \ B$) and the
superconductor, respectively.
Explicitly
$
H^{\alpha}_{N}
=
\sum_{\vect{k}\sigma} \xi_{\vect{k}}
c^{\alpha \dag}_{\vect{k}\sigma} c^{\alpha}_{\vect{k}\sigma}$
and
$
H_{S} =
\sum_{\vect{q}\sigma} \zeta_{\vect{q}} d^{\dag }_{\vect{q}\sigma}
d_{\vect{q}\sigma} + \sum_{\vect{q}} [ \Delta d^{\dag
}_{\vect{q}\downarrow} d^{\dag }_{-\vect{q} \uparrow} + \Delta^{\star}
d_{-\vect{q}\uparrow} d_{\vect{q} \downarrow} ]
$
with $\Delta$ the superconducting gap.
The operators
$c^{\alpha}_{\vect{k}\sigma}$ and $ d_{\vect{q}\sigma}$ are
the destruction operators for electrons of spin $\sigma$ in the normal metal
$\alpha$ and in the superconductor, respectively.
The indices $\vect{k}$ and $\vect{q}$ together with the spin indicate
quantum numbers labelling the eigenstates in the disconnected leads.
The tunneling part of the Hamiltonian is given by:
\begin{equation}
\label{HT}
H^{\alpha}_{T}
=
\sum_{\vect{k}\vect{q}\sigma} [
t^{\alpha}_{\vect{k}\vect{q}} c^{\alpha \dag}_{\vect{k}\sigma}
d_{\vect{q}\sigma}  + t^{\alpha \star}_{\vect{k}\vect{q}}
d^{\dag}_{\vect{q}\sigma}  c^{\alpha}_{\vect{k}\sigma}] \ ,
\end{equation}
where
$ t^{\alpha}_{\vect{k}\vect{q}}
=
\int \upd \vect{r} \upd
\vect{r}' \ t^{\alpha}(\vect{r},\vect{r}')
\psi_{\vect{k}}(\vect{r})  \psi^{\star}_{\vect{q}}(\vect{r}')$
and
$t^{\alpha}(\vect{r},\vect{r}')$ is the quantum amplitude for an electron
to tunnel from position $\vect{r}$ in the normal lead $\alpha$, to
position $\vect{r}'$ in the superconductor.
We introduce the electrostatic potential by the standard
gauge transformation $\xi_{\vect{k}} \rightarrow
\xi^{\alpha}_{\vect{k}}=\xi_{\vect{k}}+e V_{\alpha}$.

We are interested in the subgap regime defined by
$|eV_{A}|$, $|eV_{B}|$, $k_{B}T \ll \Delta $.
Then, no excitation can be created in the superconductor over times
larger than $\hbar / \Delta$.
Under these conditions, in lowest order of perturbation theory, only
two different processes contribute to charge transport:
Andreev reflection and  elastic cotunneling.
In both cases the corresponding quantum amplitude can be
obtained using second order perturbation theory \cite{Hekking}.
Specifically, for the Andreev process, one can calculate
the quantum amplitude for destroying a Cooper pair in
the superconductor and create two electrons, one
in the lead $\alpha$ with spin $\uparrow$ and the other in
lead $\beta$ with spin $\downarrow$.
The result is:
\begin{equation}
\label{Akk}
A^{\alpha \beta}_{\vect{k} \vect{p}}
=
\sum_{\vect{q}}
t^{\alpha }_{\vect{k} \vect{q}} t^{\beta }_{\vect{p} -\vect{q}}
u_{\vect{q}} v_{\vect{q}}
\left[
 \frac{1}{\xi^{\alpha}_{\vect{k}} +
   E_{\vect{q}} } +
 \frac{1}{\xi^{\beta}_{\vect{p}} + E_{\vect{q}}}
\right] \ ,
\end{equation}
where
$ u_{\vect{q}}^{2}
  =
  1 - v_{\vect{q}}^{2} =
  (1+ \zeta_{\vect{q}}/E_{\vect{q}})/2
$
and
$
E_{\vect{q}} = \sqrt{\Delta^{2}+\zeta^{2}_{\vect{q}}}$
is the superconducting spectrum.
The amplitudes $A^{AA}$ and $A^{BB}$ correspond to the injection of
both electrons in the same lead, while $A^{AB}$ and $A^{BA}$ give the
amplitude of the crossed Andreev reflection.

In a similar way one can calculate the amplitude
for elastic cotunneling $(B,\sigma) \rightarrow (A,\sigma)$:
\begin{equation}
  \label{Tkk}
  T_{\vect{k} \vect{p}}
  =
  \sum_{\vect{q}} t^{A}_{\vect{k} \vect{q}}
  t^{B \star}_{\vect{p} \vect{q}}
  \left[
    \frac{v^{2}_{\vect{q}}  }{E_{\vect{q}} +\xi^{\alpha}_{\vect{k}} }
    -
    \frac{u^{2}_{\vect{q}}  }{ E_{\vect{q}} -\xi^{\beta}_{\vect{p}}}
\right]
\,.
\end{equation}

It is convenient to write an effective Hamiltonian that
takes into account these two processes \cite{Pistolesi}:
\begin{equation}
\label{HTeff} H^{\textrm{eff}}= H^{A}_{N} +H^{B}_{N}+
\sum_{\alpha, \beta} [ J_{\alpha \beta } + h.c. ] + T + T^{\dag},
\end{equation}
where
$ J_{\alpha \beta }  = \sum_{\vect{k} \vect{p}}
A^{\alpha\beta}_{\vect{k} \vect{p}}   c^{\alpha
\dag}_{\vect{k}\uparrow} c^{\beta \dag}_{\vect{p}\downarrow} $
and
$ T = \sum_{\vect{k} \vect{p} \sigma} T_{\vect{k} \vect{p}} c^{A
\dag}_{\vect{k}\sigma} c^{B }_{\vect{p}\sigma} $.
This new Hamiltonian is equivalent to the initial one  at
low energies and it allows to obtain current and
noise straightforwardly.

\section{Currents and cross-correlation}

Current operator in each normal lead is defined, as
usual, by the time derivative of the particle number operator
$N_{\alpha} =
 \sum_{\vect{k} \sigma} c^{\alpha  \dag}_{\vect{k}\sigma}
c^{\alpha}_{\vect{k}\sigma} $ : $  I_{\alpha}= - e \ \upd
N_{\alpha} / \upd t = -(e i / \hbar)  [H^{\textrm{eff}},N_{\alpha}
]$
(sign conventions are defined in Fig. \ref{layout}).
We thus obtain:
\begin{equation}
\label{I} I_{A}= \frac{i e}{\hbar} [ 2 J_{AA} + J_{AB}+ J_{BA} + T
]  + h.c. \quad ; \quad I_{B}= \frac{i e}{\hbar} [ 2 J_{BB} +
J_{AB}+ J_{BA} - T ] + h.c. \,.
\end{equation}

It is now possible to evaluate the current in both normal leads
and their zero-frequency self- and cross-correlation defined as:
\begin{equation}
\label{Sabdef}
S_{\alpha \beta}
=
\int^{+ \infty}_{-\infty}
\!\!\!\! dt
\,
\qav{[ \delta I_{\alpha}(t) , \delta I_{\beta}(0)]_{+}}
\, ,
\end{equation}
where $\delta I_{\alpha}(t) = I_{\alpha}(t)  - \qav{I_{\alpha}}$,
$\alpha$ and $\beta$ take the values $A$ or $B$,
and the brackets denote quantum and statistical averaging.
One can then apply standard time-dependent perturbation theory
to the lowest non vanishing order ($\sim t^4$).
We obtain:
\begin{eqnarray}
\label{meanIA}
\qav{I_{A}} & =  & \frac{2 e}{\hbar^2}
\left[
  \left( N^{AA}_{\rightarrow} - N^{AA}_{\leftarrow} \right) +
  \left( N^{AB}_{\rightarrow} - N^{AB}_{\leftarrow} \right) +
  \left( N^{EC}_{\rightarrow} - N^{EC}_{\leftarrow} \right)
\right]
\ ,
\\
\label{meanSAA}
 S_{AA} & =
& \frac{4 e^2}{\hbar^2}
\left[
  2 \left( N^{AA}_{\rightarrow} + N^{AA}_{\leftarrow} \right) +
  \left( N^{AB}_{\rightarrow} + N^{AB}_{\leftarrow} \right) +
  \left( N^{EC}_{\rightarrow} + N^{EC}_{\leftarrow} \right)
\right]
\,
\\
\label{meanSAB}
 S_{AB} & =
& \frac{4 e^2}{\hbar^2}
 \left[
 \left(N^{AB}_{\rightarrow} + N^{AB}_{\leftarrow}\right) -
 \left(N^{EC}_{\rightarrow} + N^{EC}_{\leftarrow}\right)
 \right]
\,,
\end{eqnarray}
with
\begin{eqnarray}
     N^{\alpha \beta}_{ \rightleftarrows} & =  & 2 \pi \hbar
    \sum_{\vect{k}\vect{p}} |A^{\alpha \beta}_{\vect{k} \vect{p}}|^{2}
     H^{\textrm{And}}_{ \rightleftarrows} (\xi_{\vect{k}}, \xi_{\vect{p}}) \delta ( \xi_{\vect{k}}+
     \xi_{\vect{p}} + e V_{\alpha} + e V_{\beta}  )\ , \\
       N^{EC}_{ \rightleftarrows} & = & 2 \pi \hbar
    \sum_{\vect{k}\vect{p}} |T_{\vect{k} \vect{p}}|^{2}
     H^{\textrm{cot}}_{ \rightleftarrows} (\xi_{\vect{k}}, \xi_{\vect{p}}) \delta ( \xi_{\vect{k}}-
     \xi_{\vect{p}} + e V_{A} - e V_{B}  )\ ,
\end{eqnarray}
and
$  H^{\textrm{And}}_{ \rightarrow} (\xi_{\vect{k}},
\xi_{\vect{p}}) = f(\xi_{\vect{k}}) f(\xi_{\vect{p}})$ ,
$H^{\textrm{And}}_{ \leftarrow} (\xi_{\vect{k}}, \xi_{\vect{p}}) =
[1-f(\xi_{\vect{k}})] [ 1- f(\xi_{\vect{p}})]$ , $
H^{\textrm{cot}}_{ \rightarrow} (\xi_{\vect{k}}, \xi_{\vect{p}}) =
f(\xi_{\vect{k}}) [1- f(\xi_{\vect{p}})] = H^{\textrm{cot}}_{
\leftarrow} (\xi_{\vect{p}}, \xi_{\vect{k}})$,
where
$f(\xi_{k}) = \qav{ c^{\alpha \dag}_{\vect{k} \sigma}
c^{\alpha}_{\vect{k} \sigma} }$ is the Fermi function.

Eq. (\ref{meanIA}) for the current agrees with the result
obtained by  Falci {\it et al.} in Ref.  \cite{Falci}.
As mentioned above the average current has
three components: the direct Andreev ($N^{AA}$ or $N^{BB}$),
the crossed Andreev ($N^{AB}$),
and the elastic cotunneling ($N^{EC}$) currents.
The direct Andreev term is instead absent in the
cross-correlation given by Eq. (\ref{meanSAB}):
cross-correlation is thus a direct measure of CAR and EC.
It is convenient to introduce
$S^{CAR} = 4 e^2 / \hbar^2 (N^{AB}_{\rightarrow} +   N^{AB}_{\leftarrow} )$
and
$S^{EC} = 4 e^2 / \hbar^2 (N^{EC}_{\rightarrow} +   N^{EC}_{\leftarrow} )$
in terms of which we have:
$S_{AB} = S^{CAR} - S^{EC}$.
Since $N^{AB}_\rightleftarrows$ and $N^{EC}_\rightleftarrows$
are always positive, this expression implies
that CAR and EC processes contribute to $S_{AB}$
with positive and negative sign, respectively.
A simple interpretation of this fact is that
CAR implies instantaneous currents of the {\em same}
sign in both leads, while EC
implies instantaneous currents of {\em opposite} signs.

To obtain more quantitative predictions we follow the
procedure of Refs. \cite{Hekking,Pistolesi}.
The main result is given in Eq. (\ref{Sab}) below, with
the amplitudes given in Eqs. (\ref{Abal}) and (\ref{Apointcontact})
for the ballistic and diffusive regimes, respectively.

Introducing $U = (V_{A}+V_{B})/2$ and $G_Q=2e^2/h$, $S^{CAR}$ can be rewritten as
\begin{eqnarray}
  \label{Si}
  S^{CAR}
  =
  2 G_Q \, \int \upd \varepsilon
  \  A(\varepsilon)
  \  \left[
          f(\varepsilon+eU)\left(1-f(\varepsilon-eU)\right)
         +f(\varepsilon-eU)\left(1-f(\varepsilon+eU)\right)
     \right]
  \ ,
\end{eqnarray}
with
$F(\zeta; \xi_{A} , \xi_{B} ) = 2\pi u(\zeta) v(\zeta) \{
(E(\zeta)+eV_A-\xi_A)^{-1} + (E(\zeta)+ e V_{B}+\xi_{B})^{-1}\} $,
and
\begin{equation}
 \label{Adefinition}
  A(\varepsilon)
  =
  \int \upd \zeta \upd \zeta' F(\zeta;
  \varepsilon -e U , \varepsilon + e U ) F(\zeta';
  \varepsilon - e U , \varepsilon + e U )
  \,
  \Xi ( \varepsilon- e U , \varepsilon + e U ; \zeta,\zeta') \ .
\end{equation}
The function $\Xi$ reads:
\begin{eqnarray}
  \Xi(\xi_{A},\xi_{B};\zeta,\zeta')
  &=&
  \int_{\cal{V}_{A}}
  \! \! \! d \vect{r}_1   d \vect{r}_2
  \int_{\cal{V}_{B}}
  \! \! \!  d \vect{r}_3   d \vect{r}_4
  \int_{\cal{V}_{S}}
  \! \! \!  d \vect{r}'_1 \dots   d \vect{r}'_4 \,
  {t^A}^*( \vect{r}_1,\vect{r}'_1)
  {t^A}( \vect{r}_2,\vect{r}'_2)
  {t^B}^*( \vect{r}_3,\vect{r}'_3)
  {t^B}( \vect{r}_4,\vect{r}'_4)
 \nonumber \\
&& K_{\xi_{A}}  (  \vect{r}_1, \vect{r}_2 )
 K_{\xi_{B}}  (  \vect{r}_3, \vect{r}_4 )
 K_{\zeta}  (  \vect{r}'_2, \vect{r}'_3 )
 K_{\zeta'}  (  \vect{r}'_4, \vect{r}'_1 )
\end{eqnarray}
where
$ K_{\xi} ( \vect{r}_{1}, \vect{r}_{2} ) =\sum_{\vect{k}}
\delta(\xi-\xi_{\vect{k}}) \psi_{\vect{k}}(\vect{r}_{1})
\psi^{\star}_{\vect{k}}(\vect{r}_{2})$
is the single particle
spectral function, ${\cal V}_A$, ${\cal V}_B$, ${\cal V}_S$
indicate the volumes of leads A and B, and of the superconductor.

We assume that tunneling only occurs with constant amplitude and
between neighboring points lying on the junctions surface
${\cal S}_\alpha$.
Following Ref. \cite{AverinNazarov,Hekking} one can then trade the
$t$, $K_{\xi_{A}}$, and $K_{\xi_{B}}$ dependence for
the normal-state tunnel conductances per unit surface, $g^A$ and $g^B$,
(plus the density of states of the superconductor, $\nu_S$):
\begin{equation}
  \label{Xi}
  \Xi(\xi_{A},\xi_{B}; \zeta,\zeta')
  \simeq
  \Xi(\zeta,\zeta')
  =
  \frac{\hbar^2 g^{A} g^{B}}{ 16 \pi^{2}
    e^{4} \nu^{2}_{S} }
  \int_{{\cal S}_A} \!\!\! \upd^{2}\vect{r}_{A}
  \int_{{\cal S}_B} \!\!\! \upd^{2} \vect{r}_{B}
  \,
  K_{\zeta} (  \vect{r}_{A}, \vect{r}_{B} )
  K_{\zeta'}(  \vect{r}_{B}, \vect{r}_{A} )
\ .
\end{equation}
Fermi functions in Eq. (\ref{Si}) restrict the range of
variation of $|\varepsilon|$ to be at most of the
order of $\textrm{max}(k_{B}T,|eV_{A}|,|eV_{B}|) \ll \Delta$.
One can then safely discard voltage and $\varepsilon$ dependence
of $F$ in Eq. (\ref{Adefinition}).
A similar procedure can be applied to $S^{EC}$ as well.
We finally obtain:
\begin{equation}
  \label{Sab}
  S_{AB}
  =
  2e G_Q \,
  \left[
  (  V_{A} +  V_{B} )
  \coth \left( \frac{ e V_{A} + e V_{B}} {2 k_{B}T}\right)
  A^{CAR}
  - (  V_{A} -  V_{B} )
  \coth \left( \frac{ e V_{A} - e V_{B}}{2 k_{B}T}\right)
  A^{EC}
  \right]
\end{equation}
with
$
 A^{i}
 = \int \upd \zeta \upd \zeta' \,
 \Phi^{i}(\zeta )
 \Phi^{i}(\zeta') \,
 \Xi (\zeta,\zeta')$,
where $i=\{ CAR , EC \}$,
$\Phi^{CAR}(\zeta)= 2 \pi \, \Delta/(\Delta^2+\zeta^2)$
and
$\Phi^{EC}(\zeta)=2 \pi\, \zeta/(\Delta^2+\zeta^2)$.
Eq.~(\ref{Sab}) singles out
the voltage and temperature dependence of the
cross-correlation.
The spatial dependence due to the coherent propagation of electrons in
the superconductor between the two junctions
is isolated in the amplitudes
$A^{CAR}$ and $A^{EC}$.
The same amplitudes appear in the expressions for the current
obtained from (\ref{meanIA}):
\begin{equation}
  \label{finalcurrent}
  \qav{I_{A}}
  =
  I^o_{A}(V_A)  +
    G_Q\,
    (V_A+V_B) A^{CAR}
    +
    G_Q\,
    (V_A-V_B) A^{EC}
    \ .
\end{equation}
Here $I^o_{A}(V_A)$ is the direct Andreev contribution discussed
by Hekking and Nazarov\cite{Hekking} giving rise to the non linear
$I$-$V$ characteristic.
The second and third term on the rhs of Eq. (\ref{finalcurrent})
give instead the CAR and EC contribution to the current and they
have been discussed in Refs.~\cite{Byers,Deutscher,Loss,Falci}.
For completeness we give the current auto-correlation (noise):
\begin{equation}
 S_{AA}
 =
 4 e I^o_A(V_A) {\rm coth} \left( eV_A/k_B T\right)+
 S^{CAR}+S^{EC}
\ .
\end{equation}
The first term of the rhs was obtained in Ref. \cite{Pistolesi}
and dominates the noise.

Comparing Eqs. (\ref{Sab}) and (\ref{finalcurrent}) the
advantage of measuring the low-temperature cross correlation over
the mean current is apparent.
Indeed, in $S_{AB}$ for $T=0$ there is no direct term.
This allows to measure directly $A^{CAR}$ or $A^{EC}$ by
setting $V_A=V_B$ or $V_A=-V_B$, respectively.
In contrast, the same procedure would not work
by measuring the current.
As we will see below, $A^{CAR} \approx A^{EC} = A$,
thus the current becomes
$
  \qav{I_{A}}
  =
  I^o_{A}(V_A)  +
    2 G_Q \, A\, V_A
$, with no residual dependence on $V_B$.
The interesting $A$ dependence is thus hidden by the
direct term $I^o$.
In order to raise the amplitude degeneracy some authors
\cite{Deutscher,Falci,Melin,Taddei, Chtchelkatchev, Feinberg} proposed to perform
the experiment with spin-polarized normal leads, and measure
the drag current $I_A(V_A=0,V_B)$.
This is not necessary if $S^{AB}$ is measured.
A last remark on the temperature dependence is in order.
For temperatures $k_B T \gg |eV_A|,|eV_B|$ cross-correlation
is completely suppressed.

We discuss now briefly the amplitude in the two interesting
regimes: ballistic and diffusive.
In the ballistic regime one finds \cite{Loss,Falci}:
\begin{eqnarray}
    \label{Abal}
    \left(
    \begin{array}{c}
     A^{CAR} \\
     A^{EC}
    \end{array}
    \right)
    = \frac{ g^{A}  g^{B} }{G_Q^2  }
    \int_{{\cal S}_A}     \int_{{\cal S}_B}
      \upd^2 \vect{r}_{A}
      \upd^2 \vect{r}_{B}
    \left(
    \begin{array}{c}
    \sin^{2}(k_{F}|\vect{r}_{A} - \vect{r}_{B}|) \\
    \cos^{2}(k_{F}|\vect{r}_{A} - \vect{r}_{B}|)
    \end{array}
    \right)
    \frac{e^{-2 |\vect{r}_{A} - \vect{r}_{B}|/ \xi_{b} }}{
    (k_{F}|\vect{r}_{A} - \vect{r}_{B}|)^{2}} \ ,
\end{eqnarray}
where $\xi_b=\hbar v_F/\Delta$ is the superconducting
coherence length and $v_F$, $k_F$ the Fermi velocity
and momentum.
In multichannel junctions, the trigonometric functions
in (\ref{Abal}) are averaged out, giving the same numerical
factor. Thus $ A^{CAR} \approx A^{EC} \equiv A_b $.
For junctions with a typical size much smaller than the distance, $R$,
between the two contacts, the spatial
dependence is given by $A_b \sim e^{-2 R/ \xi_{b}}/(k_{F} R)^{2}$.
The exponential decay is related to the characteristic size of
the Cooper pairs in the superconducting lead.
The algebraic prefactor further reduces the intensity of the effect on
the scale of the Fermi wavelength
as discussed in Refs. \cite{Loss,Falci,Prada}.
Cross contributions are expected to be small in
ballistic systems.

\begin{figure}
   \centerline{\includegraphics[scale=0.8]{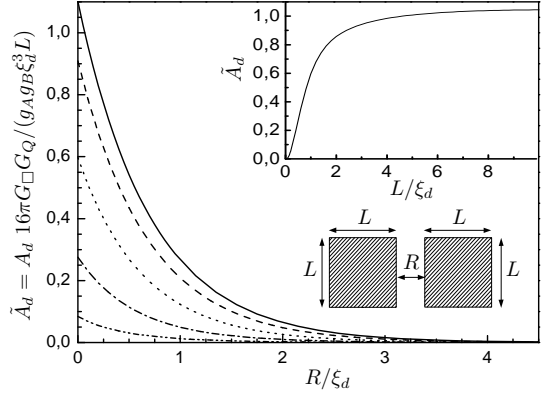}}
   \caption{
Amplitude $A_d$ for two square contacts on a diffusive superconductive
film. From top to bottom: $L/\xi=\infty$, 5, 2, 1, 0.5.
Dependence on the distance $R$ for different side lengths $L$.
Inset: dependence on the side length for $R=0$.
}
    \label{magnitude}
\end{figure}

In diffusive superconductors the coherence length $\xi_{d} =
\sqrt{\hbar D /\Delta} $ ($D$ is the diffusion constant) and the
distance between the two junctions largely exceed the elastic mean
free path $l_{e}$.
Impurity averaging is performed on (\ref{Xi}) \cite{Chtchelkatchev,Feinberg}
within the usual approximation $k_{F}l_{e} \ll 1$:
\begin{equation}
  \qav{K_{\zeta}
    (  \vect{r}_{A}, \vect{r}_{B} ) K_{\zeta'}
    (\vect{r}_{B}, \vect{r}_{A} )}_{\textrm{imp}}
  =
  \frac{\nu_{S}}{2 \pi}
  \left[
  P_{\zeta - \zeta'}(\vect{r}_{A}, \vect{r}_{B} ) +
  P_{\zeta' - \zeta} ( \vect{r}_{B}, \vect{r}_{A} )
  \right] \ .
\end{equation}
The Cooperon $ P_{\epsilon} $ satisfies the diffusion equation
in  the superconducting lead:
\begin{equation}
-\hbar D \Delta_{\vect{r}_{1}}P_{\epsilon}-i \epsilon P_{\epsilon}
=  \delta^{3} ( \vect{r}_{1} - \vect{r}_{2} ) \ .
\end{equation}
Then, we find
\begin{eqnarray}
\label{Apointcontact}
A_d
\equiv A^{CAR}
= A^{EC}
= \frac
{\hbar^2 g^{A} g^{B}}
{ 32 \pi e^{4} \nu_{S}}
\int \upd^2 \vect{r}_{A} \upd^2
\vect{r}_{B} P_{2 i \Delta}(|\vect{r}_{A} - \vect{r}_{B}|) \ .
\end{eqnarray}
If the superconducting electrode is a thin film with a thickness
$d \ll \xi_{d}$ (characterized  by its sheet conductance
$G_\Box= 2 e^2 \nu_S D d$ in the normal phase),
we can use the picture of two-dimensional
electron diffusion.
Then we find $P_{2 i \Delta}(R)=(1/\pi \hbar D d)K_0(\sqrt{2}R/\xi_d)$,
where $K_{0}$ is the modified Bessel function of the second order.
In a thicker film, $P_{2 i \Delta}(R)=(1/4\pi \hbar D R)e^{\sqrt{2}R/\xi_d}$.
We calculate explicitly the value of $A_d$ when the normal leads
are square films of side $L$ deposited at distance $R$
over a superconducting film.
The results are presented in Fig.~\ref{magnitude}.
As expected, the distance dependence is essentially determined by
$P_{2i\Delta}(R)$.
The size dependence is more interesting.
For large values of $L/\xi_d$, $A_d(R=0)$ grows linearly with $L$,
this is due to the fact that the main contribution comes from a small
stripe of width $\xi_d$ and length $L$ in each contact.
This can be used experimentally to increase the signal,
there are no draw backs in increasing the lateral side of the
contacts.
For small contact sizes $L\ll \xi_d$, the whole contact contributes
with the same value giving $A_d \sim L^4 P_{2i\Delta}(R)$.

\section{Conclusion}

We calculated the current-current cross-correlation in multiterminal
hybrid structures consisting of two normal metallic leads attached to
a superconductor through tunnel barriers. We found that varying the
voltage biases between the normal leads and the superconductor allows
to probe separately two non-local mechanisms for the charge transfer
in such systems. We evaluated this effect for realistic devices and we
found that it may be observable in diffusive structures, provided that
the two normal arms are deposited at a distance not larger than few
tens of nanometers (of the order of the superconducting coherence length in the
diffusive regime).
In view of the recent observation of the doubled shot noise in a
superconducting/normal metal tunnel junction \cite{Lefloch} we think
that measuring such effects is within the reach of present technology.

\acknowledgments
We thank D. Feinberg for pointing out the advantage of special
geometries of contacts.
G.B, F.P., and F.W.J.H. acknowledge financial support
from CNRS/ATIP-JC 2002 and Institut Universitaire de France.
MH acknowledges financial support from the
ACI-JC no 2036 from the French Minstry of Research.


\begin{thebibliography}{0}

\bibitem{Andreev}
    \Name{Andreev A.~F.}
    \REVIEW{Sov. Phys. JETP}{19}{1964}{1228}
    (\Name{Andreev A.~F.}
    \REVIEW{Zh. Eksp. Teor. Fiz.}{46}{1964}{1823}).

\bibitem{Byers}
    \Name{Byers J.~M. \and Flatt\'{e} M.~E.}
    \REVIEW{Phys. Rev. Lett.}{74}{1995}{306}.

\bibitem{Deutscher}
    \Name{Deutscher G. \and Feinberg D.}
    \REVIEW{Appl. Phys. Lett.}{76}{2000}{487}.

\bibitem{Loss}
    \Name{Recher P., Sukhoroukov E.~V. \and Loss D.}
    \REVIEW{Phys. Rev. B}{63}{2001}{165314}.

\bibitem{Falci}
    \Name{Falci G., Feinberg D. \and Hekking F.~W.~J.}
    \REVIEW{Europhys. Lett.}{54}{2001}{255}.

\bibitem{AverinNazarov}
    \Name{Averin D.V. \and Nazarov Yu. V.}
    \REVIEW{Phys. Rev. Lett.}{65}{1990}{2446}.

\bibitem{Blanter}
  For a recent review see
  \Name{Blanter Y.~M., \and B\"{u}ttiker M.}
  \REVIEW{Phys. Rep.}{336}{2000}{1}.

\bibitem{Datta}
    \Name{Anantram M.~P. \and Datta S.}
    \REVIEW{Phys. Rev. B}{53}{1996}{16 390}.

\bibitem{Martin}
    \Name{Martin T.}
    \REVIEW{Phys. Lett. A}{220}{1996}{137}.

\bibitem{TorresMartin}
    \Name{Torr\`es J. \and Martin T.}
    \REVIEW{Eur. Phys. J. B}{12}{1999}{319}.

\bibitem{TML}
    \Name{Torr\`es J., Martin T. \and Lesovik G.~B.}
    \REVIEW{Phys. Rev. B}{63}{2001}{134517}.

\bibitem{LMB}
    \Name{Lesovik G.~B., Martin T. \and Blatter G.}
    \REVIEW{Eur. Phys. J. B}{24}{2001}{287}.

\bibitem{Shechter}
   \Name{Shechter M., Imry J. \and Levinson Y.}
    \REVIEW{Phys. Rev. B}{24}{2001}{287}.

\bibitem{Taddei}
    \Name{Taddei F. \and Fazio R.}
    \REVIEW{Phys. Rev. B}{65}{2002}{134522}.

\bibitem{Borlin}
    \Name{B\"orlin J., Belzig W. \and Bruder C.}
    \REVIEW{Phys. Rev. Lett.}{88}{2002}{197001}.

\bibitem{Samuelsson}
  \Name{Samuelson P. \and B\"uttiker M.}
  \REVIEW{Phys. Rev. Lett.}{89}{2002}{046601}.

\bibitem{Samuelsson2}
  \Name{Samuelson P. \and B\"uttiker M.}
  \REVIEW{Phys. Rev. B}{66}{2002}{201306}.

\bibitem{Book}
 For general reviews see also
  \Book{Quantum noise in mesoscopic systems}
  \Editor{Nazarov Yu.~V.}
  \Publ{Kluwer, Dordrecht}
  \Year{2002}.

\bibitem{Hekking}
    \Name{Hekking F.~W.~J.\and Nazarov Y.~V.}
    \REVIEW{Phys. Rev. Lett.}{71}{1994}{1625} and
    \REVIEW{Phys. Rev. B}{49}{1994}{6847}.
\bibitem{Pistolesi}
  \Name{Pistolesi F., Bignon G. \and Hekking F. W. J.}
  ArXiv:cond-mat/0303165.

\bibitem{Melin}
    \Name{Melin R. \and Feinberg D.}
    \REVIEW{Eur. Phys. J. B.}{26}{2002}{101}.



\bibitem{Chtchelkatchev}
  \Name{Chtchelkatchev N.~M. \and Burmistrov I.~S.}
  ArXiv:cond-mat/0303014;
  \Name{Chtchelkatchev N.~M. \and Mar'enko M.}
  ArXiv:cond-mat/0306552.
\bibitem{Feinberg}
  \Name{Feinberg D.}
  ArXiv:cond-mat/0307099.
\bibitem{Prada}
  \Name{Prada E. \and Sols F.}
  ArXiv:cond-mat/0307500.
\bibitem{Lefloch}
    \Name{Lefloch F., Hoffmann C., Sanquer M. \and Quirion D.}
    \REVIEW{Phys. Rev. Lett.}{88}{2002}{197001}.
\end{thebibliography}
\end{document}